\documentclass[aps,prd,nofootinbib,longbibliography,11pt]{revtex4-2}
\usepackage{etoolbox}
\usepackage{dcolumn}
\usepackage{amsmath,amssymb,amsfonts,mathtools}
\usepackage{mathrsfs}
\usepackage{graphicx}
\usepackage[colorlinks=true,urlcolor=blue,citecolor=red,linkcolor=blue]{hyperref}
\usepackage{natbib}
\usepackage{wrapfig}
\usepackage{flushend}

\begin{document}
\title{Minimal length scale correction in the noise of gravitons}
\author{Sukanta Bhattacharyya}
\email{sukanta706@gmail.com}
\affiliation{Department of Physics, West Bengal State University, Barasat, Kolkata 700126, India}
\author{Soham Sen}
\email{sensohomhary@gmail.com}
\affiliation{Department of Astrophysics and High Energy Physics, S. N. Bose National Centre for Basic Sciences, JD Block, Sector-III, Salt Lake City, Kolkata-700 106, India}
\author{Sunandan Gangopadhyay}
\email{sunandan.gangopadhyay@gmail.com}
\affiliation{Department of Astrophysics and High Energy Physics, S. N. Bose National Centre for Basic Sciences, JD Block, Sector-III, Salt Lake City, Kolkata-700 106, India}
\title{{\bf{Resonant detectors of gravitational wave in the linear and quadratic generalized uncertainty principle framework}}}
\begin{abstract}
\noindent In this work, we consider a resonant bar detector of gravitational waves in the generalized uncertainty principle (GUP) framework with linear and quadratic momentum uncertainties. The phonon modes in these detectors vibrate due to the interaction with the incoming gravitational wave. In this uncertainty principle framework, we calculate the resonant frequencies and transition rates induced by the incoming gravitational waves on these detectors. We observe that the energy eigenstates and the eigenvalues get modified by the GUP parameters. We also observe non-vanishing transition probabilities between two adjacent energy levels due to the existence of the linear order momentum correction in the generalized uncertainty relation which was not present in the quadratic GUP analysis [\href{http://dx.doi.org/10.1088/1361-6382/abac45}{Class. Quantum Grav. 37 (2020) 195006}]. We finally obtain bounds on the dimensionless GUP parameters using the form of the transition rates obtained during this analysis.

\end{abstract}
\maketitle

\section{Introduction}
\noindent
With the advent of quantum mechanics and general relativity at the beginning of the twentieth century, the primary focus of the scientific community shifted towards the discovery of a combined description of the behaviour of the physical theories which is effective at a large length scale as well as in the subatomic domain.  Quantum mechanics gives an almost perfect theory at the subatomic length scale and the general theory of relativity is extremely accurate at the astrophysical length scale. Now, if one tries to write down a quantum theory of gravity, it is mandatory to probe general relativity in the subatomic length scale. In quantum mechanics, we consider the geometry of spacetime as a flat stationary background, while in general relativity energy or matter can create deformation in the flat structure of spacetime resulting in a curved spacetime geometry. Therefore, to explore the universe near the Planck scale, we need a complete quantum theory of gravity that will give rise to quantum mechanics and general relativity in their respective domains. All the attempts to quantize gravity, like loop quantum gravity \cite{lqg1,lqg2}, string theory \cite{ad,kk}, noncommutative geometry \cite{gi}, and some gedanken experiments in quantum gravity phenomenology unanimously indicates the existence of an observer-independent minimal length scale. All such calculations suggest that this minimum length scale should be identified with the Planck length which implies $l_{pl} \sim 10^{-35}$ m. 

\noindent The ideal way to canonically quantize a theory is to raise the phase space variables to operator status and then implement a commutation relation among the two canonically conjugate variables. Such commutation relation among the phase space variables leads eventually to the Heisenberg's uncertainty principle. One way to implement an observer-independent minimal length scale in a theory is to modify Heisenberg's uncertainty principle which is in general known as the generalized uncertainty principle (GUP) in the literature. Interestingly, among other consequences of this fundamental hypothesis of this Planck-length, the modification of the Heisenberg uncertainty principle (HUP) to the generalized uncertainty principle (GUP) has gained a deep interest in a plethora of investigations in high energy theoretical physics. In this development, Mead first proposed  \cite{mead} that gravity affects the uncertainty principle and leads to the GUP. Several investigations into black hole physics, M-theory \cite{yon,yon1,yon2}, and the nonperturbative formulation of open string field theory \cite{wit1,hor} indicate that the modification of HUP is proportional to the quadratic term in momentum. However, it is important to note that, for such theories with modified uncertainty relations, the classical and quantum side of the approach results in some inconsistencies. A rigorous Hamiltonian and Lagrangian analysis with classical and quantum systems has been done in \cite{Bosso}. When the commutator between the position and momentum variables gets modified due to the inclusion of minimal length into the theory, it is no more admissible to call the position and momentum variables as conjugate variables. To avoid this terminological issue, we have restricted ourselves from using the term ``conjugate" corresponding to the modified phase space variables of the system in this analysis. In a different study \cite{Bosso2}, it is observed that the physical content of a theory resonates via the commutator of the physical observables. In this paper, it is finally argued that the minimal length scale is physical and cannot be removed by resorting to a different commutation relation. Recently in a study \cite{OTMApple}, we have used the noise of gravitons induced on a freely falling particle (under the effect of Earth's gravity) to obtain quantum gravity corrections to the Heisenberg uncertainty principle and obtained quadratic order corrections in momentum with explicit dependence on the three fundamental constants ($\hbar$, $G$, and $c$). In the Planck mass limit, it reduces to the well-known form of the generalized uncertainty principle (GUP).  However, doubly special relativity theory suggests that this modification should contain a term that should be linear in momentum. Now, combining both possibilities, one can find a further modification to the GUP, which is also known as the linear and quadratic generalized uncertainty principle (LQGUP). This idea has been first introduced in \cite{sd2}.

\noindent Hence, one gets the modified GUP between the position $Q_i$ and the momentum $P_j$ incorporating both the contributions from the linear and quadratic order terms in momentum as \cite{sd1},
\begin{equation}\label{gupvp}
\Delta Q_i \Delta P_i\geq~ \frac{\hbar}{2}\left[1-\alpha \left\langle P+\frac{P_i P_i}{P}\right\rangle-(\alpha^2-\beta)\left[(\Delta P)^2+{\langle P\rangle}^2\right]-(\alpha^2-2\beta)\left[({\Delta P_i})^2+{\langle P_i\rangle}^2\right]\right],
\end{equation}
\noindent  where ${{P}}^2\equiv{\vert\vec{{P}}\vert}^2= \eta_{ij}P^{i}P^{j}$; $i,j=1,2,3$. The parameters $\alpha$ and $\beta$ are positive and independent of $\Delta Q$ and $\Delta P$ which generate the traces of the GUP. These parameters are defined as $\alpha=\alpha_{0}/(M_{Pl}c)$ and $\beta=\beta_{0}/(M_{Pl}c)^2$ where $M_{Pl}$ is the Planck mass and $c$ is the speed of light in free space. The definitions of GUP parameters
immediately show that the dimensions of $\alpha$ and $\beta$ are $(\text{momentum})^{-1}$ and $(\text{momentum})^{-2}$ respectively. Now the modified Heisenberg algebra following the uncertainty relation (\ref{gupvp}) is as follows
\begin{equation}\label{gupu}
[Q_i,P_j]=i\hbar\left[\delta_{ij}-\alpha\left(\delta_{ij}P+\dfrac{P_i P_j}{P}\right)+\beta\left(\delta_{ij}P^2+2P_i	P_j\right)-\alpha^2(\delta_{ij}P^2+P_i P_j)\right]~.
\end{equation}
\noindent The existence of GUP has been immensely investigated at the level of black hole thermodynamics  \cite{mmg,fsc,rjadler,rd,rb,sg,fs1,sbbh}, various quantum systems like a particle in a box, Landau-level, harmonic oscillator  \cite{sd1,sd2}, \cite{id, pb, pb11} and so on. Theoretical analysis with this GUP framework in the path integral representation  \cite{sp,sb,sb1} has also been thoroughly investigated to gain more insight into the quantum gravity phenomenology. In a recent study, the noise spectrum of a typical optomechanical setup has been analyzed in the  LQGUP framework \cite{ss}. In \cite{QFTToyModel}, a quantum harmonic oscillator is studied non-perturbatively, showing that a linear correction to the commutation relation between position and momentum cannot admit any ladder operators. Despite all these theoretical investigations in the GUP framework, this Planck scale phenomenon poses the question of how one goes about experimentally capturing its signature.  Before investigating any relics of GUP in any experimental setup, it is important to note that the upper bound on the dimensionless GUP parameter $\beta_0$ reported in recent studies \cite{sd1, fs1, bawaj, zf, bushev, fs}, corresponds to the intermediate length scale $l_{im} = \sqrt{\beta_0}l_{pl} \sim 10^{-23} \text{m}$.

\noindent Resonant bar detectors \cite{jw1,jw2} of gravitational waves (gravitational wave) have gained potential interest following the breakthrough observation of the gravitational wave by LIGO \cite{bp} and VIRGO recently, and the subsequent observations of numerous additional gravitational wave sources \cite{bp1,bp2,bp3}. These observations open up a new epoch where one can hope to explore the signature of quantum gravity which is generally of the order of the Planck length scale. Therefore, researchers are interested in investigating whether any signature of quantum gravity effect lies in the gravitational wave detector scenario \cite{GW-detection_status}. Recently the highly sensitive detectors can detect the variations $\Delta L$ of the bar length $L \sim 1 \text{m}$ with a sensitivity of the order of $\frac{\Delta L}{L} \sim 10^{-19}$. Now this scale of length variation will be more precise as the sensitivity of the detectors increases, which leads to a better understanding of the universe.  The main motivation behind the investigation of resonant bar detectors is that they are technically much more straightforward to construct than the interferometer-based `L' shaped detectors like LIGO or VIRGO. Again it is also theoretically very prudent to investigate such detectors as they involve the interaction of gravitational waves with elastic matter. The interaction of gravitational waves with simple matter systems causes tiny vibrations which are many orders smaller than the size of an atomic nucleus. This tiny vibration can be modeled as a quantum mechanical forced harmonic oscillator. Therefore, it is possible to simply write down the resonant detectors of gravitational waves as a gravitational wave-harmonic oscillator interaction model. The possibility of detecting these Planck length relics in the gravitational wave detection technique motivates us to do a rigorous investigation of the quantum mechanical responses of the gravitational wave detectors to the gravitational wave fluctuations in noncommutativity and GUP framework \cite{sbf,sb2,sg1,sg22,sg3,sg4,sg5,OTM}. Recently, in\cite{bushev}, we have seen that the measurements of resonant frequencies of a mechanical oscillator bear the signature of GUP. In a different way, an optomechanical scheme has been introduced in \cite{pikov}, which was further developed in \cite{bosso1, kumar}, for realizing the presence of GUP in  experimental scenarios. 

\noindent Motivated by the above discussion, in this paper, we present the quantum mechanical effects of the gravitational wave detectors in the LQGUP framework. Here, we consider the LQGUP which contains both the liner and quadratic order contributions in momentum. In our work, we have mainly investigated the transition probabilities using the Fermi-Golden rule for several gravitational wave templates. Remarkably, due to the presence of the linear order correction in the momentum in the modified uncertainty relation, we observed few transitions from the ground state of the harmonic oscillator which was absent in the case of the modified uncertainty relation with quadratic order correction in the momentum only. If one can attenuate all other noise contributions that can lead to such a transition, then these transitions can confirm the existence of LQGUP.

\noindent The construction of the paper goes as follows. In section 2, we construct the Hamiltonian of the gravitational wave-harmonic oscillator system in the LQGUP framework (\ref{gupu}). Then we calculate the formal perturbative solutions of that system in section 3. Next, we apply the time-independent perturbation theory to get the perturbed eigenstates of the one-dimensional harmonic oscillator with the new perturbed energy eigenvalues. After that, we use the time-dependent part of the Hamiltonian to calculate the transition between the states of the harmonic oscillator caused by the gravitational wave. In section 4, we use the various gravitational wave waveforms to calculate the corresponding transition probabilities. We finally conclude in section 5.
\section{The harmonic oscillator-gravitational wave interaction model}
\noindent  For the background spacetime of our system we have considered small time-dependent fluctuations over the flat Minkowski background as
\begin{equation}\label{1.1}
g_{\mu\nu}=\eta_{\mu\nu}+h_{\mu\nu}
\end{equation}
with $h_{\mu\nu}$ denoting these fluctuations (specifically gravitational wave fluctuations) over the flat Minkowski background. Here $\eta_{\mu\nu}=\text{diag}\{-1,1,1,1\}$ in eq.(\ref{1.1}).

\noindent To investigate the response of the gravitational wave detectors interacting with the incoming gravitational wave in the LQGUP framework given by eq.(\ref{gupu}) which contains both the linear and quadratic contributions in momentum, we first need to write down the Hamiltonian of the harmonic oscillator system representing the elastic matter interacting with a gravitational wave. The representation of the dynamics of the elastic matter by a harmonic oscillator follows from the fact that the fundamental mode of the elastic bar is described by a harmonic oscillator \cite{Magg} . In order to detect such tiny vibrations in the elastic matter due to gravitational wave, one needs to make use of linear amplifiers to detect an acoustic oscillation of the fundamental mode of the bar.  This acoustic oscillation of the fundamental mode of the bar can be best described by the oscillation of a single phonon mode \cite{Magg}. Inspite of the bar being a macroscopic object (weighing of the order of a ton), the oscillation of the bar is detected which is not explained properly by a classical treatment. Therefore, in this case one considers the collective effect of the number of excited-phonons to describe the oscillation of the bar.

\noindent Now the dynamics of the harmonic oscillator-gravitational wave system can be properly framed by the geodesic deviation equation for a two-dimensional harmonic oscillator of mass $m$ and intrinsic frequency $\omega$ in a proper detector frame as
\begin{equation}
m \ddot{{Q}} ^{j}= - m{R^j}_{0,k0} {Q}^{k} - m \omega^{2} Q^{j}~; ~j=1,2~.
\label{e5}
\end{equation}
The Riemann curvature tensor in eq.(\ref{e5}) in terms of the gravitational fluctuations can be represented as ${R^j}_{0,k0} = - \frac{d \Gamma^j_{0k}}{d t}  = -\ddot{h}_{jk}/2$. Here we consider the plane of the elastic matter in the $x-y$ plane and the direction of the propagation vector of the gravitational wave to be in the $z$ direction. The dot denotes the derivative with respect to the coordinate time of the proper detector frame. It is the same as its proper time to first order in the metric perturbation and ${Q}^{j}$ is the proper distance of the pendulum from the origin. In particular, $Q_j$ is best described by the displacement of the individual phonon modes from their equilibrium positions. Here the spatial velocities are nonrelativistic and the coordinate $|{Q}^{j}|$ is much smaller than the reduced wavelength $\frac{\lambda}{2\pi}$ of the gravitational wave.
Now using the transverse-traceless (TT) gauge condition one can remove the unphysical degrees of freedom and get only two relevant components, namely the $\times$ and $+$ polarizations of the gravitational wave arising in the analytical form of the Riemann curvature tensor ${R^j}_{0,k0} = -\ddot{h}_{jk}/2$. 
Thus, the gravitational wave interaction gives rise to a time-dependent piece in eq.(\ref{e5}) \footnote{This TT gauge condition makes the spatial part of the gravitational wave to be unity ( $e^{i \vec{k}.\vec{x}} \approx 1$) all over the detector site in case of the plane-wave expansion of gravitational wave.}. Therefore, $h_{jk}$ containing the polarization information reads
\begin{equation}
h_{jk} \left(t\right) = 2f (t)\left(\varepsilon_{\times}\sigma^1_{jk} + \varepsilon_{+}\sigma^3_{jk}\right)
\label{e13}
\end{equation}
\noindent where $\sigma^1$ and $\sigma^3$ are the Pauli spin matrices, $2f$ is the amplitude of the gravitational wave, and $\left( \varepsilon_{\times}, \varepsilon_{+} \right)$ are the two possible polarization states of the gravitational wave satisfying the condition $\varepsilon_{\times}^2+\varepsilon_{+}^2 = 1$ for all $t$. Note that for the linearly polarized gravitational wave, the frequency  $\Omega$ of the gravitational wave is contained in the time-dependent amplitude $2f(t)$ and the time-dependent polarization states $\left( \varepsilon_{\times} \left(t \right), \varepsilon_{+} \left( t \right) \right)$ contain the frequency $\Omega$ for the circularly polarized gravitational wave.

\noindent In the next step, we find the Lagrangian describing the interaction of the elastic matter system with the gravitational wave from the geodesic deviation eq.(\ref{e5}) as
\begin{equation}
{\cal L} = \frac{1}{2} m\dot {Q}_{j}\dot {Q}^{j} - m{\Gamma^j}_{0k}
\dot {Q}_{j} {Q}^{k}  - \frac{1}{2} m \omega^{2}  Q_{j}Q^{j}
\label{e8}
\end{equation}
  where $Q_{j}Q^{j}= \sum_{j=1}^{3}~Q^{j}Q_{j}\simeq \sum_{j=1}^{3}~Q_{j}Q_{j}$.

\noindent Hence, computing the momentum ${P}^{j} = m\dot {Q}^{j} - m \Gamma^j_{0k} {Q}^{k}$ corresponding to ${Q}_{j}$ from the above Lagrangian, one immediately can write down the Hamiltonian of the gravitational wave-harmonic oscillator system as
\begin{equation}
{H} = \frac{\delta_{ij}}{2m}\left({P}^{i} + m \Gamma^i_{~0k} {Q}^{k}\right)\left({P}^{j} + m \Gamma^j_{~0l} {Q}^{l}\right) + \frac{1}{2} m \omega^{2} Q_{j}Q^j~ .
\label{e9}
\end{equation}
\noindent Now, we want to explore whether the gravitational wave detectors can testify to any signature of GUP incorporating both the contributions from the linear and quadratic order corrections in momentum. To do this, first, we build the quantum mechanical description of the system. At first, we replace the phase-space variables $\left( Q_{j}, P_{j} \right)$ by the operators  $\left( \hat{Q}_{j}, {\hat P}_{j} \right)$. Then we use the map between the modified operators $(\hat{Q}_j, \hat{P}_j)$ in terms of the usual canonically conjugate operators $(\hat{q}_{j},\hat{p}_{j})$ up to first order in $\beta$ obeying eq.(\ref{gupu}) as
\begin{equation}
	\hat{Q}_{i}=\hat{q}_{i}~,~~\hat{P}_{i}=\hat{p}_{i}(1-\alpha \hat{p}_{i} +\beta \hat{p}^{2}) 
	\label{GUPR}
\end{equation}
where $\hat{q}_{i},~\hat{p}_{j}$ satisfy the usual canonical commutation relation $[\hat{q}_{i}, \hat{p}_{j}]=i\hbar\delta_{ij}$ and $\hat{p}^2=\sum_{i=1}^3\hat{p}_{i}\hat{p}_{i}$. Here $q_i$ and $p_j$ are canonically conjugate variables as they satisfy the usual commutation relation.

\noindent 
Before proceeding further, note that a typical bar of gravitational wave-detector has a length $L= 3 \text{ m}$ and radius $R= 30$ cm. Therefore, it is straightforward to approximate the gravitational wave-harmonic oscillator system as a one-dimensional system as a whole. It is quite simple to understand that because of this one-dimensional treatment, the cross-polarization term does not survive anymore. The Hamiltonian describing the system up to order $(\alpha^2, \beta) $ reads {\footnote{For notational simplicity, we use $\hat p_{j} \equiv p_j$ and $\hat q_{j}\equiv q_j$. For the one dimensional case we have used $p_1\equiv p$ and $q_1\equiv q$.}} 
\begin{equation}
H=\left(\frac{p^2}{2m}+\frac{1}{2} m \omega^2 q^2 \right)+\frac{1}{2}\Gamma^1_{01} \left(pq+qp\right)-\frac{\alpha p^3}{m} -\frac{\alpha}{2}\Gamma^1_{01} \left(p^2 q+q p^2\right)+\frac{\alpha^2p^4}{2m}+\frac{\beta p^4}{m} +\frac{\beta}{2}\Gamma^1_{01}\left(p^3q+qp^3\right)
\label{4}
\end{equation}
where $p_1=p$ and $q_1=q$.
\noindent It can be seen that the above Hamiltonian is Hermitian. Hence we break the Hamiltonian (\ref{4}) as
\begin{eqnarray}
H=H_{0}+H_{1}+H_{2}
\label{5b}
\end{eqnarray}
where the analytical forms of $H_0$, $H_1$, and $H_2$ are given by
\begin{align}
H_0&=\frac{p^2}{2m}+\frac{1}{2}m\omega^2q^2 \label{51a}\\
H_1&=-\frac{\alpha}{m}p^3+\frac{\gamma}{m}p^4~,~\gamma=\left(\frac{\alpha^2}{2}+\beta\right)  \label{52a}\\
H_2&=\frac{1}{2} \Gamma^1_{01}\left(pq+qp\right)-\frac{\alpha}{2}\Gamma^1_{~01}\left(p^2q+qp^2\right)+\frac{\beta}{2}\Gamma^1_{01}\left(p^3q+qp^3\right)~.
\label{5a}
\end{align}
The above expressions show that we can treat $H_1$ and $H_2$ as perturbative Hamiltonians which arise due to the gravitational wave interaction and the underlying GUP framework. The GUP parameters $\alpha $ and $\beta$ give rise to the time-independent part of the Hamiltonian denoted by $H_1$.    
Therefore with this background in place, to analyze the quantum mechanical behavior of the gravitational wave detector under the LQGUP framework, first we calculate the shift in the energy eigenvalues and the modified energy eigenstates of the harmonic oscillator due to $H_1$. Up to this point, no interaction terms arise due to the incoming gravitational wave. Now the time-dependent Hamiltonian $H_2$ arises due to the interaction of the gravitational wave with the harmonic oscillator system. The time-dependent part of the Hamiltonian is responsible for the transitions between the perturbed energy eigenstates. Our aim in this paper is to calculate these transition amplitudes and the corresponding transition probabilities as in the future if such transition probabilities can be measured, it will indicate the existence of an observer-independent minimal length scale in nature. Again we can easily see that $H_2$ also contains terms that are linear in $\alpha$ and $\beta$. Such terms may be more significant in our theoretical setup.          

\noindent Now, the Hamiltonians in eq.(s)(\ref{51a},\ref{52a},\ref{5a}) can be recast in terms of the operators $(a^{ \dagger}, a)$ as 
\begin{align}
H_0=&\hbar \omega\left(a^\dagger a +\frac{1}{2}\right) \label{13a}\\
H_{1}=&-i\frac{\alpha}{m}\left(\frac{\hbar m \omega}{2}\right)^{\frac{3}{2}}\left[a^3-a^2a^\dagger-aa^\dagger a+aa^{\dagger2}-a^\dagger a^2+a^\dagger a a^\dagger+a^{\dagger2} a-a^{\dagger3}\right]\nonumber\\& +\frac{\gamma}{m}\left(\frac{\hbar m \omega}{2}\right)^2\bigr[a^4-a^3a^\dagger-a^2a^\dagger a+a^2a^{\dagger2}-aa^\dagger a^2+(aa^\dagger)^2+aa^{\dagger2} a-a a^{\dagger3} -a^\dagger a^3+a^\dagger a^2 a^\dagger \nonumber\\&+(a^\dagger a)^2 -a^\dagger a a^{\dagger2} +a^{\dagger2} a^2-a^{\dagger2} a a^\dagger -a^{\dagger3} a+a^{\dagger4}\bigr]\label{14b} \\
H_{2} = &i \hbar \dot h_{11}\left[-\frac{1}{2}\left(a^2-a^{\dagger2}\right)+\alpha \hbar \left(\frac{\hbar m \omega }{2}\right) \left[a^3-aa^\dagger a -a^\dagger a a^\dagger+a^{\dagger3} \right]\right. \nonumber\\ & \left.+\frac{\beta \hbar m \omega}{4}\bigr(a^4-a^2 a^\dagger a-aa^\dagger a^2+a a^{\dagger2} a
-a^\dagger a^2 a^\dagger+a^\dagger a a^{\dagger2} +a^{\dagger2} a a^\dagger -a^{\dagger4}\bigr)\right] ~.
\label{14}
\end{align}
where $q$ and $p$ in terms of the raising and lowering operators are given as
\begin{align}
	q&=\left(\frac{\hbar}{2 m \omega}\right)^\frac{1}{2}\left(a+a^\dagger\right)\label{6}~,\\
	p&=-i\left(\frac{ m \hbar\omega}{2}\right)^\frac{1}{2}\left(a-a^\dagger\right)\label{6a}~.	
\end{align}
 
\noindent It is important to note that in the above equation, the operators $a$ and $a^\dagger$ act on the eigenstates of the Hamiltonian $H_0$\footnote{A way of working with ladder operators corresponding to the full system (any general system) has been shown in \cite{Bosso4}.}. We shall make use of these relations in the next section to obtain the perturbative energy levels and then find the transition probabilities between them.


\section{Perturbed energy levels and transition amplitudes}
\noindent In this section, we proceed to calculate the perturbed eigenstates due to time-independent Hamiltonian $H_1$ in eq.(\ref{14b}). Using time-independent perturbation theory, the perturbed $n$ th energy eigenstate reads
\begin{eqnarray}
|n \rangle^\beta &=& |n \rangle	-i\frac{\alpha}{m \hbar \omega} \left(\frac{\hbar m\omega}{2}\right)^\frac{3}{2} \left[\frac{\sqrt{n(n-1)(n-2)}}{3}|n-3 \rangle - 3 n \sqrt{n}|n-1 \rangle-3 (n+1)  \right. \nonumber\\ &\times& \left.\sqrt{(n+1)}|n+1 \rangle+ \frac{\sqrt{(n+1)(n+2)(n+3)}}{3} |n+3 \rangle\right] + \gamma\hbar m\omega\left[\frac{(2n+3)\sqrt{(n+1)(n+2)}}{4}\right. \nonumber\\&\times&|n+2 \rangle- \frac{(2n-1)\sqrt{n(n-1)}}{4}|n-2 \rangle  \left. + \frac{\sqrt{n(n-1)(n-2)(n-3)}}{16}|n-4\rangle\right.\nonumber\\& -&\left. \frac{\sqrt{(n+1)(n+2)(n+3)(n+4)}}{16} |n+4 \rangle\right]
\label{tipes}
\end{eqnarray}
with energy eigenvalues
\begin{eqnarray}
E_n &=& \left(n+\frac{1}{2}\right) \hbar \omega \left[1+ m \omega \hbar \gamma \frac{3(2n^2+2n+1)}{2(2n+1)}\right]	~.
\label{env}
\end{eqnarray} 

\noindent These are the general forms of the eigenstates and eigenvalues of a one-dimensional harmonic oscillator in the presence of GUP containing both linear and quadratic contributions in momentum. In eq.(\ref{tipes}), the left-hand side is the corrected eigenstate of the time-independent part of the full Hamiltonian $H_0+H_1$ whereas the $|n\rangle$ states in the right-hand side are the eigenstates of the free Hamiltonian $H_0$ such that $a|n\rangle=\sqrt{n}|n-1\rangle$ and $a^\dagger|n\rangle=\sqrt{n+1}|n+1\rangle$. Here, $a$ and $a^\dagger$ do not lower or raise $|n\rangle^\beta$ directly as they are not the raising and lowering operators of the full Hamiltonian $H$.

\noindent We now calculate the transitions between the modified states of the harmonic oscillator induced by the incoming gravitational wave which is mathematically taken care of by the time-dependent Hamiltonian $H_2(t)$ in eq.(\ref{5a}). To the lowest order approximation in time-dependent perturbation theory, the probability amplitude of transition from an initial state $|i\rangle$ to a final state $|f \rangle$, ($i\neq f$), due to a perturbation {\bf{$H_2(t)$}} is given by \cite{kurt}
\begin{eqnarray}
C_{i \rightarrow f}(t\rightarrow \infty)  =  -\frac{i}{\hbar} \int_{-\infty}^{t\rightarrow +\infty} dt' F \left( t' \right) e^{\frac{i}{\hbar}(E_f -E_i)t'} \langle \Psi_f | \hat{Q}|\Psi_i \rangle
\label{probamp}
\end{eqnarray}
where $H_2(t)=F(t)\hat{Q}$ with $F(t)=\dot h_{11} $, and $\hat{Q}$ being given by (hat has been omitted while writing the expression)
\begin{eqnarray} 
	Q &=& i \hbar \left[-\frac{1}{2}\left(aa-a^\dagger a^\dagger\right)+\alpha \hbar \left(\frac{\hbar m \omega }{2}\right) \left[aaa-aa^\dagger a -a^\dagger a a^\dagger+a^\dagger a^\dagger a^\dagger \right] \right. \nonumber\\ && \left. +\frac{\beta \hbar m \omega}{4}\left(aaaa-aa a^\dagger a-aa^\dagger aa+a a^\dagger a^\dagger a
	-a^\dagger aa a^\dagger+a^\dagger a a^\dagger a^\dagger +a^\dagger a^\dagger a a^\dagger -a^\dagger a^\dagger a^\dagger a^\dagger\right)\right] ~.
	\label{14a}
\end{eqnarray}

\noindent Using eq.(\ref{14a}) 
in eq.(\ref{probamp}), we can write down all of the possible transition amplitudes from the perturbed ground state as 
\begin{eqnarray}
C_{0^\beta \rightarrow 1^\beta } &=& T_{01} \int_{-\infty}^{t\rightarrow +\infty} dt'~~ \dot h_{11} ~~e^{i(1+3 \gamma m \omega \hbar )\omega t'}\label{10a}\\ 
C_{0^\beta \rightarrow 2^\beta } &=& T_{02} \int_{-\infty}^{t\rightarrow +\infty} dt'~~ \dot h_{11}~~e^{i(2+9 \gamma m \omega \hbar )\omega t'} \label{10b}\\ 
C_{0^\beta \rightarrow 3^\beta} &=& T_{03} \int_{-\infty}^{t\rightarrow +\infty} dt'~~ \dot h_{11} e^{i(3+18 \gamma m \omega \hbar )\omega t'} \label{10c}\\ 
C_{0^\beta \rightarrow 4^\beta} &=& T_{04} \int_{-\infty}^{t\rightarrow +\infty} dt'~~ \dot h_{11}~~e^{i(4+30 \gamma m \omega \hbar)\omega t'}
\label{10}
\end{eqnarray}
where $T_{01},~T_{02},~T_{03}$, and $T_{04}$ are given by
\begin{eqnarray}
T_{01} &=& 	-i\alpha \sqrt{\frac{m\hbar \omega}{2}} \nonumber\\
T_{02} &=& \frac{1}{2\sqrt{2}} \left[1+\frac{17}{2}\alpha^2 m \omega \hbar + \frac{9}{4} \beta m \omega \hbar\right] \nonumber\\
T_{03} &=& - i \alpha \sqrt{3 m \omega \hbar} \nonumber\\
T_{04}&=& -\frac{\sqrt{6}}{4} \left[\frac{29}{12} \alpha^2 m \omega \hbar +3 \beta m \omega \hbar\right]~.
\label{conts}
\end{eqnarray}
Now looking at the above transition amplitudes, we can make the following observations. In the limit $\alpha \rightarrow 0$ and $\beta \rightarrow 0$, all the transition amplitudes become zero except $C_{0^\beta \rightarrow 2^\beta }$. Hence for a one-dimensional harmonic oscillator interacting with gravitational wave under the standard HUP framework, only $ 0^\beta \rightarrow 2^\beta$ transition will occur. Due to the presence of the GUP containing only quadratic term in momentum (i.e., $\alpha=0$), we get another transition $0^\beta\rightarrow 4^\beta$ in addition to the previous one $( 0^\beta \rightarrow 2^\beta )$ with different amplitudes \cite{sb}. In this paper, we consider the LQGUP which contains both the contributions in linear and quadratic terms in momentum. This consideration reveals that there are transitions between $0^\beta \rightarrow 1^\beta$, $0^\beta \rightarrow 2^\beta$, $0^\beta \rightarrow 3^\beta$ and $0^\beta \rightarrow 4^\beta$ with different amplitudes. Moreover, note that $T_{01}$ and $T_{03}$ arises due to presence of only $\alpha$ and $T_{02}$ and $T_{04}$ bear the contributions of both $\alpha$ and $\beta$. Therefore, looking at the possible experimental outcomes \cite{sg4,sg5,sb} of these detectors in a future generation of resonant bar detectors, one can see that resonant detector of the gravitational wave is not only a reliable candidate to probe such Planckian effects but also be able to predict the mathematical structure of such Planckian modifications.
This is one of the most important results in this paper. It is important to note that such transitions can occur also if certain external noise contribution is there which may lead to such transitions between adjacent states as well. Hence, one needs to get rid of all the additional noise contributions to truly detect the existence of LQGUP in nature.

\noindent Now eq.(s)(\ref{10a},\ref{10b},\ref{10c},\ref{10}) denoting the transition amplitudes indicate that the presence of the LQGUP can be checked by measuring the corresponding transition probabilities from the relation
\begin{eqnarray}\label{prob}
P_{i\rightarrow f}=  |C_{i\rightarrow f}|^{2}.
\label{tp}
\end{eqnarray}
\noindent In the next section, we shall calculate the transition probabilities for different types of incoming gravitational waves.


\section{Transition probabilities for different types of gravitational wave templates}
\noindent We now investigate the analytical forms of the transition probabilities for the harmonic oscillator interacting with different gravitational wave templates. Such different gravitational wave templates can indeed be generated in nature in complex astronomical events. At first, we start with the simplest possible gravitational wave template which is a linearly polarized gravitational wave.  

\subsection{Periodic linearly polarized gravitational wave}
\noindent In this subsection, we consider a periodic gravitational wave with linear polarization. The analytical form of the gravitational perturbation indicating a linearly polarized gravitational wave reads
\begin{equation}
h_{jk} \left(t\right) = 2f_{0} \cos{\Omega t} \left(\varepsilon_{\times}\sigma^1_{jk} + \varepsilon_{+}\sigma^3_{jk}\right)~.
\label{lin_pol}
\end{equation}
The amplitude $h_{jk}(t)$ varies periodically with a single frequency $\Omega$. In order to obtain the transition probabilities we shall make use of the form of $h_{jk}$ from eq.(\ref{lin_pol}) and substitute it back in eq.(s)(\ref{10a}-\ref{10}) and obtain the transition amplitudes. Using eq.(\ref{prob}), it is now possible to write down all of the physically possible transition probabilities from the obtained transition amplitudes
(where we got rid of the transitions for which $\omega$ is negative as they are not possible physically)
\begin{equation}\label{1.30}
\begin{split}
P_{0^\beta\rightarrow 1^\beta} &= \left(2 \pi f_0 \Omega |T_{01}| \epsilon_+\right)^2 \times \left[\delta\left( \omega\left(1 +3 \gamma m \omega \hbar  \right)- \Omega \right) \right]^2 \\	
P_{0^\beta\rightarrow 2^\beta} &= \left(2 \pi f_0 \Omega |T_{02}| \epsilon_+\right)^2 \times \left[\delta\left( \omega\left(2 +9 \gamma m \omega \hbar \right)- \Omega \right) \right]^2 \\
P_{0^\beta\rightarrow 3^\beta} &=  \left(2 \pi f_0 \Omega |T_{03}| \epsilon_+\right)^2 \times \left[\delta \left( \omega \left(3 +18 \gamma m \omega \hbar \right)-\Omega \right) \right]^2~\\
P_{0^\beta\rightarrow 4^\beta} &=  \left(2 \pi f_0 \Omega |T_{04}| \epsilon_+\right)^2 \times \left[\delta \left( \omega \left(4 +30 \gamma m \omega \hbar \right)-\Omega \right) \right]^2~.
\end{split}
\end{equation}
\noindent When a real experimental observation is done then the observational time is finite, therefore we can regularize the Dirac delta function as $\delta(\omega)=\left[ \int_{-\frac{T}{2}}^{\frac{T}{2}} dt~~  e^{i \omega t} \right] = T$. It is now possible to write down all of the possible transition rates (after getting rid of the unphysical part in the transition probabilities)
\begin{equation}
\begin{split}
\lim\limits_{T \rightarrow \infty} \frac{1}{T}P_{0^\beta\rightarrow 1^\beta} &=    \left(2 \pi f_0 \Omega |T_{01}| \epsilon_+\right)^2 \times \left[ \delta\left( \omega\left(1 +3 \gamma m \omega \hbar  \right)- \Omega \right) \right]
 \\
\lim\limits_{T \rightarrow \infty} \frac{1}{T}P_{0^\beta\rightarrow 2^\beta} &=   \left(2 \pi f_0 \Omega |T_{02}| \epsilon_+\right)^2 \times \left[ \delta\left( \omega\left(2 +9 \gamma m \omega \hbar \right)- \Omega \right) \right]
 \\
\lim\limits_{T \rightarrow \infty} \frac{1}{T}P_{0^\beta\rightarrow 3^\beta} &=    \left(2 \pi f_0 \Omega |T_{03}| \epsilon_+\right)^2 \times \left[ \delta \left( \omega \left(3 +18 \gamma m \omega \hbar \right)-\Omega \right) \right] \\
\lim\limits_{T \rightarrow \infty} \frac{1}{T}P_{0^\beta\rightarrow 4^\beta} &=   \left(2 \pi f_0 \Omega |T_{04}| \epsilon_+\right)^2 \times \left[ \delta \left( \omega \left(4 +30 \gamma m \omega \hbar \right)-\Omega \right) \right]~.
\label{trlpgw}
\end{split}
\end{equation}
\noindent From the above expressions we can easily find the resonant frequencies and amplitudes of the transitions between the different modified states. Note that for an ordinary harmonic oscillator, there is only one transition $P_{0\rightarrow 2}$ at resonant frequency $\Omega=2 \omega$. Now in this paper, we consider the general form of the LQGUP containing both the linear and quadratic contributions in momentum and get four transitions namely $P_{0^\beta\rightarrow 1^\beta}$, $P_{0^\beta\rightarrow 2^\beta}$, $P_{0^\beta\rightarrow 3^\beta}$ and $P_{0^\beta\rightarrow 4^\beta}$.  Among them  $P_{0^\beta\rightarrow 1^\beta}$ and $P_{0^\beta\rightarrow 3^\beta}$ take place at frequencies $\Omega= \omega(1+3 \gamma m \omega \hbar)$ and $\Omega = \omega(3+18 \gamma m \omega \hbar)$. Here the amplitudes of the transitions are proportional to $\alpha^2$. Therefore, only the linear contributions of momentum to the GUP are seen in these two transitions. On the other hand the transitions $P_{0^\beta\rightarrow 2^\beta}$ and $P_{0^\beta\rightarrow 4^\beta}$ occur due to both the GUP parameters $\alpha $ and $\beta$. These show resonances at frequencies $\Omega= \omega(2+9\gamma m \omega \hbar)$ and $\Omega = \omega (4+30 \gamma m \omega \hbar)$ retain the terms linear and quadratic of both the parameters $\alpha$ and $\beta$.  

\noindent 
Hence, this theoretical model not only detects signatures of GUP but may also prove to be a reliable candidate to validate the form of the proposed GUP structure in the future.

\subsection{Circularly polarized gravitational wave}
\noindent The next template that we will discuss is the periodic gravitational wave signal with circular polarization which can be expressed as
\begin{equation}
h_{jk} \left( t \right) = 2f_{0} \left[\varepsilon_{\times} \left( t \right) \sigma^1_{jk} + \varepsilon_{+}\left( t \right) \sigma^3_{jk}\right] 
\label{cir_pol}
\end{equation}
where $\varepsilon_{+} \left( t \right)  = \cos \Omega t $ and $\varepsilon_{\times} \left( t \right)  = \sin \Omega t $ with $\Omega$ denoting the gravitational wave frequency. Following the same approach as in the linear polarization case,  the transition rates corresponding to the gravitational wave template in eq.(\ref{cir_pol}) reads
 \begin{equation}
 \begin{split}
	\lim\limits_{T \rightarrow \infty} \frac{1}{T}P_{0^\beta\rightarrow 1^\beta} &=   \left(2\pi f_0\Omega |T_{01}|  \right)^2 \times \left[ \delta\left( \omega\left(1 +3 \gamma m \omega \hbar  \right)- \Omega \right) \right] \\
	\lim\limits_{T \rightarrow \infty} \frac{1}{T}P_{0^\beta\rightarrow 2^\beta} &=   \left(2 \pi f_0 \Omega |T_{02}| \right)^2 \times \left[ \delta\left( \omega\left(2 +9 \gamma m \omega \hbar \right)- \Omega \right) \right]\\
	\lim\limits_{T \rightarrow \infty} \frac{1}{T}P_{0^\beta\rightarrow 3^\beta} &=    \left(2 \pi f_0 \Omega |T_{03}| \right)^2 \times \left[ \delta \left( \omega \left(3 +18 \gamma m \omega \hbar \right)-\Omega \right) \right] \\
	\lim\limits_{T \rightarrow \infty} \frac{1}{T}P_{0^\beta\rightarrow 4^\beta} &=  \left(2 \pi f_0 \Omega |T_{04}| \right)^2 \times \left[ \delta \left( \omega \left(4 +30 \gamma m \omega \hbar \right)-\Omega \right) \right]~.
	\label{trcpgw}
\end{split}
\end{equation}
Eq.(\ref{trcpgw}) have striking similarities with that of the transition rates obtained in the case of the linearly polarized gravitational wave in eq.(\ref{trlpgw}) indicating that the circularly polarized gravitational waves can also be good candidates in signatures of LQGUP in future generation of resonant bar detectors.

 
\subsection{Aperiodic linearly polarized gravitational wave: gravitational wave burst}
\noindent Next we consider the gravitational wave with aperiodic signals. Such aperiodic signals are generated from the in-spiral neutron stars or black hole binaries in general. When the two black holes or neutron stars merge with each other, they emit gravitational wave signals with huge amounts of energy. The duration of such signals is very small and they lie in the time range of $10^{-3}$ sec$< \tau_g< 1$ sec. Such emission of gravitational waves with high energy are commonly known as bursts. The analytical way of expressing such bursts is given as
\begin{eqnarray}
h_{jk} \left(t\right) = 2f_{0} g \left( t \right) \left(\varepsilon_{\times}\sigma^1_{jk} + \varepsilon_{+}\sigma^3_{jk}\right).
\label{lin_pol_burst}
\end{eqnarray}
As the duration of such signals must be very small, the smooth function $g \left( t \right)$ must go to zero very rapidly for time $|t| > \tau_{{\rm g}}$\footnote{Note that $\tau_g \sim \frac{1}{f_{max}}$. Here $f_{max}$ is the maximum value of a broad range of continuum spectrum of frequency.}. Let us take a Gaussian form for the function $g(t)$
\begin{equation}
g \left(t\right) = e^{- t^{2}/ \tau_{g}^{2}}.
\label{burst_waveform_Gaussian}
\end{equation}
As the gravitational wave burst is very short-lived, it contains a wide range of frequency values \cite{sb2}.  
It is now possible to write down the gravitation wave burst in terms of its Fourier transform as
\begin{eqnarray}
h_{jk} \left(t\right) = \frac{f_{0}}{\pi} \left(\varepsilon_{\times}\sigma^1_{jk} + \varepsilon_{+}\sigma^3_{jk}\right)  \int_{-\infty}^{+\infty} \bar{g} \left( \Omega \right) e^{- i \Omega t}  d \Omega ~.
\label{lin_pol_burst_Gaussian}
\end{eqnarray}
In the above equation, we have defined $\bar{g} \left( \Omega \right)\equiv\sqrt{\pi} \tau_{g} e^{- \left( \frac{\Omega \tau_{g}}{ 2} \right)^{2}}$ . It is rather straightforward to see that $\bar{g}(\Omega)$ is the amplitude of the Fourier mode at frequency $\Omega$. 
\noindent Before writing down all of the possible transition amplitudes from the ground state to higher excited energy states, we shall explicitly calculate one of the transition amplitudes as follows.
\begin{equation}\label{transition1}
\begin{split}
C_{0^\beta\rightarrow 1^\beta}&=-i\alpha \sqrt{\frac{m\hbar\omega}{2}}\int_{-\infty}^{\infty}dt e^{\frac{i}{\hbar}(E_1-E_0)t}\dot{h}_{11}(t)\\
&=\frac{4i\alpha f_0\varepsilon_+}{\tau_g^2}\sqrt{\frac{m\hbar\omega}{2}}\int_{-\infty}^\infty dt\hspace{0.2mm} t \hspace{0.2mm}e^{-\frac{t^2}{\tau_g^2}}e^{i\omega t(1+3\gamma \hbar\omega)} \\
&=\frac{4i\alpha f_0\varepsilon_+}{\tau_g^2}\sqrt{\frac{m\hbar\omega}{2}}e^{-\frac{\omega^2\tau_g^2}{4}(1+3\gamma\hbar\omega)^2}\int_{-\infty}^\infty dt\hspace{0.2mm} t \hspace{0.2mm}e^{-\left(\frac{t}{\tau_g}-\frac{i\omega \tau_g}{2}(1+3\gamma\hbar m\omega)\right)^2}~.
\end{split}
\end{equation}
the above integral is a Gaussian integral and it is straightforward to obtain the analytical form of the transition amplitude in eq.(\ref{transition1}) as
\begin{equation}\label{transition2}
\begin{split}
C_{0^\beta\rightarrow 1^\beta}&=-2if_0\varepsilon_+T_{01}(\omega+3\gamma m\hbar\omega^2)\sqrt{\pi}\tau_ge^{-\left(\frac{\tau_g}{2}(\omega+3\gamma \hbar m \omega^2)\right)^2}\\
&=-2if_0\varepsilon_+T_{01}(\omega+3\gamma m\hbar\omega^2)\bar{g}(\omega+3\gamma m\hbar \omega^2)~.
\end{split}
\end{equation} 
Following the above procedure one can compute the transition amplitudes corresponding to other allowed transitions as
\begin{equation}
\begin{split}
C_{0^\beta\rightarrow 2^\beta} &=-2 i f_0 \varepsilon_{+} T_{02} \left(2\omega +9 \gamma m \omega^2 \hbar \right) \bar{g}(2\omega +9 \gamma m \omega \hbar) \\
C_{0^\beta\rightarrow 3^\beta} &=-2 i f_0 \varepsilon_{+} T_{03} \left(3\omega +18 \gamma m \omega^2 \hbar \right) \bar{g}(3\omega +18 \gamma m \omega^2 \hbar) \\
C_{0^\beta\rightarrow 4^\beta} &=-2 i f_0 \varepsilon_{+} T_{04} \left(4\omega +30 \gamma m \omega^2 \hbar \right) \bar{g}(4\omega +30 \gamma m \omega^2 \hbar)~. 
\label{15}
\end{split}
\end{equation}
The main difference of the transition amplitudes in eq.(s)(\ref{transition2},\ref{15}) from the earlier cases with linear and circularly polarized gravitational wave templates is that the Dirac delta functions no more appear in the calculation as a result of the time-dependent part being a Gaussian function. 
Using the expression of $\bar{g} \left( \Omega \right) $ we can obtain the form of the transition probabilities given as
\begin{equation}
\begin{split}
P_{0^\beta\rightarrow 1^\beta} &= \left(2  f_0 |T_{01}|\varepsilon_{+} \sqrt{\pi} \tau_g \left(\omega +3 \gamma m \omega^2 \hbar \right)          \right)^2 e^{-2 \left\{\frac{\omega +3 \gamma m \omega^2 \hbar}{2} \tau_g\right\}^2}\\
P_{0^\beta\rightarrow 2^\beta} &= \left( 2  f_0 |T_{02}| \varepsilon_{+} \sqrt{\pi} \tau_g \left(2\omega +9 \gamma m \omega^2 \hbar \right) \right)^2 e^{-2 \left\{\frac{2\omega +9  \gamma m \omega^2 \hbar }{2} \tau_g\right\}^2} \\
P_{0^\beta\rightarrow 3^\beta} &= \left( 2  f_0 |T_{03}| \varepsilon_{+} \sqrt{\pi} \tau_g \left(3\omega +18 \gamma m \omega^2 \hbar \right) \right)^2 e^{-2 \left\{\frac{3\omega +18  \gamma m \omega^2 \hbar }{2} \tau_g\right\}^2} \\
P_{0^\beta\rightarrow 4^\beta} &= \left( 2  f_0 |T_{04}| \varepsilon_{+} \sqrt{\pi} \tau_g \left(4\omega +30 \gamma m \omega^2 \hbar \right) \right)^2 e^{-2 \left\{\frac{4\omega +30  \gamma m \omega^2 \hbar }{2} \tau_g\right\}^2}~.
\label{12}
\end{split}
\end{equation}

\noindent The above exercises reveal a very beautiful insight into the underlying nature of the linear and quadratic generalized uncertainty principle. If we start with the basic case with Heisenberg's uncertainty relation, we observe that only one transition occurs from the ground state of the harmonic oscillator which is the $|0\rangle\rightarrow|2\rangle$ transition due to the interaction with gravitational wave. Now if one considers the quadratic GUP as has been analyzed in \cite{sbf}, an extra transition is also possible from the perturbed ground state to the fourth excited energy state. In the current work, due to the inclusion of the linear order correction in momentum uncertainty, we obtain the transition to first and third excited energy eigenstates as well.  Therefore, these results indicate a new window to probe the presence of quantum gravity effects in the resonant detector of the gravitational wave set up and if experimentally proven the possible transitions will also indicate a way to identify what kind of uncertainty relation nature obeys in a quantum gravity scenario.
\subsection{Gaussian wave packet}
\noindent The previous system is a simple case where gravitational wave bursts are considered for a scenario where gravitational waves are being generated from a binary system or merger phase. We shall further generalize the initial ansatz in the previous section to consider a general class of astronomical scenarios. In this subsection, we consider the case when the gravitational wave has the structure of a Gaussian wave packet. The function $g(t)$ in eq.(\ref{burst_waveform_Gaussian}) takes the form \cite{Magg}
\begin{equation}\label{4D.1}
g(t)=\left(\frac{2}{\pi}\right)^{\frac{1}{4}}\frac{1}{\sqrt{\tau_g}}e^{-i\Omega t}e^{-t^2/\tau_g^2}~.
\end{equation}
It is important to note that the above function is no longer dimensionless. Hence, the only way is to multiply the above function by a $\sqrt{\tau_g}$ factor. With the above Gaussian wave packet in hand, we can write down the classical gravitational fluctuation term as
\begin{equation}\label{4D.2}
h_{jk}(t)=2f_0\sqrt{\tau_g}g(t)(\varepsilon_\times\sigma^1_{jk}+\varepsilon_+\sigma^3_{jk}).
\end{equation}
Following the earlier analysis, we obtain the transition amplitude for going from $0^\beta$ state to $1^\beta$ state to be 
\begin{equation}\label{4D.3}
\begin{split}
&C_{0^\beta\rightarrow 1^\beta}\\&=-i\alpha \sqrt{\frac{m\hbar\omega}{2}}\int_{-\infty}^{\infty}dt~ e^{\frac{i}{\hbar}(E_1-E_0)t}\dot{h}_{11}(t)\\
&=\frac{4i\alpha f_0\varepsilon_+}{(2\pi)^{\frac{1}{4}}}\sqrt{m\hbar\omega}\int_{-\infty}^{\infty}\frac{dt}{\tau_g}~e^{-\frac{t^2}{\tau_g^2}}e^{i(\omega(1+3\gamma m\hbar \omega)-\Omega)t}\left(\frac{t}{\tau_g}+\frac{i}{2}\Omega\tau_g\right)\\
&=\frac{4i\alpha f_0\varepsilon_+}{(2\pi)^{\frac{1}{4}}}\sqrt{m\hbar\omega}e^{-\frac{\tau_g^2}{4}(\omega(1+3\gamma m\hbar\omega)-\Omega)^2}\int_{-\infty}^{\infty}d\left(\frac{t}{\tau_g}\right)\left(\frac{t}{\tau_g}+\frac{i}{2}\Omega\tau_g\right)e^{-\left(\frac{t}{\tau_g}+\frac{i}{2}\Omega\tau_g-\frac{i}{2}\omega\tau_g(1+3\gamma m\hbar \omega)\right)^2}\\
&=\frac{4i\alpha f_0\varepsilon_+}{(2\pi)^{\frac{1}{4}}}\sqrt{m\hbar\omega}e^{-\frac{\tau_g^2}{4}(\omega(1+3\gamma m\hbar\omega)-\Omega)^2}\left(\sqrt{\pi}\frac{i}{2}\omega\tau_g(1+3\gamma m\hbar\omega)\right)\\
&=-\alpha f_0\varepsilon_+\sqrt{2m\hbar\omega}\tau_g(2\pi)^{\frac{1}{4}}\omega(1+3\gamma m\hbar \omega)e^{-\frac{\tau_g^2}{4}(\omega(1+3\gamma m\hbar\omega)-\Omega)^2}\\
&=-2i f_0\varepsilon_+T_{01}(2\pi)^{\frac{1}{4}}\tau_g(\omega+3\gamma m\hbar \omega^2)e^{-\frac{\tau_g^2}{4}(\omega(1+3\gamma m\hbar\omega)-\Omega)^2}~.
\end{split}
\end{equation}
Similarly, the other transition amplitudes are given by
\begin{align}
C_{0^\beta\rightarrow 2^\beta}&=-2i f_0\varepsilon_+T_{02}(2\pi)^{\frac{1}{4}}\tau_g(2\omega+9\gamma m\hbar \omega^2)e^{-\frac{\tau_g^2}{4}(\omega(2+9\gamma m\hbar\omega)-\Omega)^2}\label{4D.4}\\
C_{0^\beta\rightarrow 3^\beta}&=-2i f_0\varepsilon_+T_{03}(2\pi)^{\frac{1}{4}}\tau_g(3\omega+18\gamma m\hbar \omega^2)e^{-\frac{\tau_g^2}{4}(\omega(3+18\gamma m\hbar\omega)-\Omega)^2}\label{4D.5}\\
C_{0^\beta\rightarrow 4^\beta}&=-2i f_0\varepsilon_+T_{04}(2\pi)^{\frac{1}{4}}\tau_g(4\omega+30\gamma m\hbar \omega^2)e^{-\frac{\tau_g^2}{4}(\omega(4+30\gamma m\hbar\omega)-\Omega)^2}\label{4D.6}~.
\end{align}
The Dirac-delta function in terms of a Gaussian function can be written as
\begin{equation}\label{4D.7}
\lim\limits_{\sigma\rightarrow 0}\frac{1}{\sqrt{2\pi}\sigma}e^{-\frac{(x-x_0)^2}{2\sigma^2}}=\delta(x-x_0)
\end{equation}
with $\sigma$ denoting the width of the Gaussian distribution. In principle, $\tau_g$ denotes the duration of measuring a single mode of the gravitational wave. One should therefore take $\tau_g\rightarrow\infty$ limit for a general gravitational wave measurement. Taking this limit, we can recast eq.(\ref{4D.3}) as
\begin{equation}\label{4D.8}
\begin{split}
\lim\limits_{\tau_g\rightarrow\infty}C_{0^\beta\rightarrow 1^\beta}&=\lim\limits_{\tau_g\rightarrow\infty}\left(-4if_0\varepsilon_+T_{01}\sqrt{\pi} (2\pi)^{\frac{1}{4}}(\omega+3\gamma m\hbar \omega^2)\frac{\tau_g}{2\sqrt{\pi}}
e^{-\frac{1}{2\left(\frac{\sqrt{2}}{\tau_g}\right)^2}(\omega(1+3\gamma m\hbar\omega)-\Omega)^2}\right)\\
&\simeq-4if_0\varepsilon_+T_{01}\sqrt{\pi} (2\pi)^{\frac{1}{4}}(\omega+3\gamma m\hbar \omega^2)\lim\limits_{\tau_g\rightarrow\infty}\left[\frac{\tau_g}{2\sqrt{\pi}}
e^{-\frac{1}{2\left(\frac{\sqrt{2}}{\tau_g}\right)^2}(\omega(1+3\gamma m\hbar\omega)-\Omega)^2}\right]~.
\end{split}
\end{equation}
Applying eq.(\ref{4D.7}) in eq.(\ref{4D.8}), we obtain
\begin{equation}\label{4D.9}
\begin{split}
\lim\limits_{\tau_g\rightarrow\infty} C_{0^\beta\rightarrow 1^\beta}&=-4if_0\varepsilon_+|T_{01}| (2\pi^3)^{\frac{1}{4}}(\omega+3\gamma m\hbar \omega^2)\delta\left(\omega(1+3\gamma m\hbar\omega)-\Omega\right)~.
\end{split}
\end{equation}
The associated transition probability reads in this limit
\begin{equation}\label{4D.10}
\begin{split}
P^\infty_{0^\beta\rightarrow 1^\beta}&=|\lim\limits_{\tau_g\rightarrow\infty}C_{0^\beta\rightarrow 1^\beta}|^2\\&=\sqrt{2\pi^3} \left(4f_0\varepsilon_+|T_{01}|(\omega+3\gamma m\hbar\omega^2)\right)^2(\delta(\omega(1+3\gamma m\hbar\omega)-\Omega))^2\\
&=4\sqrt{\frac{2}{\pi}} \left(2\pi f_0\Omega|T_{01}|\varepsilon_+\right)^2(\delta(\omega(1+3\gamma m\hbar\omega)-\Omega))^2
\end{split}
\end{equation}
where in the final line of the above equation we have used the property $f(x)\delta(x-x_0)=f(x_0)\delta(x-x_0)$ of the Dirac delta function.
The above result in eq.(\ref{4D.10}) is the same as that of the first result in eq.(\ref{1.30}) of the linearly polarized gravitational wave case up to a constant factor $4\sqrt{\frac{2}{\pi}}$. Similar results follow for the other three transition probabilities. This does establish a one-to-one correspondence among the linearly polarized gravitational wave with that of the Gaussian wave function case in the $\tau_g\rightarrow\infty$ limit up to some coefficient factor. This constant factor can be absorbed just by multiplying $h_{jk}(t)$ with a number $\frac{1}{2}\left(\frac{\pi}{2}\right)^{\frac{1}{4}}$ which recasts eq.(\ref{4D.2}) as 
\begin{equation}
\begin{split}
h_{jk}(t)&=\left(\frac{\pi}{2}\right)^{\frac{1}{4}}f_0\sqrt{\tau_g}g(t)(\varepsilon_\times\sigma^1_{jk}+\varepsilon_+\sigma^3_{jk})
\\&=f_0e^{-i\Omega t}e^{-t^2/\tau_g^2}(\varepsilon_\times\sigma^1_{jk}+\varepsilon_+\sigma^3_{jk})~.
\end{split}
\end{equation}
We can similarly also calculate all other transition amplitudes and probabilities in this $\tau_g\rightarrow\infty$ limit but we are not listing them as they will follow the same structure. The transition probabilities for the Gaussian wave function read
\begin{align}
P_{0^\beta\rightarrow 1^\beta}&=\left(2 f_0\varepsilon_+|T_{01}|\right)^2(2\pi)^{\frac{1}{2}}\tau_g^2(\omega+3\gamma m\hbar \omega^2)^2e^{-\frac{\tau_g^2}{2}(\omega(1+3\gamma m\hbar\omega)-\Omega)^2}\label{4D.11}\\
P_{0^\beta\rightarrow 2^\beta}&=\left(2 f_0\varepsilon_+|T_{02}|\right)^2(2\pi)^{\frac{1}{2}}\tau_g^2(2\omega+9\gamma m\hbar \omega^2)^2e^{-\frac{\tau_g^2}{2}(\omega(2+9\gamma m\hbar\omega)-\Omega)^2}\label{4D.12}\\
P_{0^\beta\rightarrow 3^\beta}&=\left(2 f_0\varepsilon_+|T_{03}|\right)^2(2\pi)^{\frac{1}{2}}\tau_g^2(3\omega+18\gamma m\hbar \omega^2)^2e^{-\frac{\tau_g^2}{2}(\omega(3+18\gamma m\hbar\omega)-\Omega)^2}\label{4D.13}\\
P_{0^\beta\rightarrow 4^\beta}&=\left(2 f_0\varepsilon_+|T_{04}|\right)^2(2\pi)^{\frac{1}{2}}\tau_g^2(4\omega+30\gamma m\hbar \omega^2)^2e^{-\frac{\tau_g^2}{2}(\omega(4+30\gamma m\hbar\omega)-\Omega)^2}\label{4D.14}~.
\end{align}
\subsection{Obtaining bounds on the dimensionless GUP parameters}
\noindent In this subsection, our primary aim is to obtain bounds on the dimensionless GUP parameter $\alpha_0$ and $\beta_0$. We shall consider only the linear polarization case to obtain the bounds on the dimensionless GUP parameter.
From eq.(\ref{trlpgw}), it is straightforward to observe that the Dirac delta functions in the transition rates are dependent on the frequency of the harmonics oscillator, gravitational wave frequency as well as the parameter $\gamma$ which is dependent on the parameters $\alpha$ and $\beta$. It is quite intuitive to claim that the GUP correction part must be smaller than the leading order $\omega$ correction. Hence, it is possible to write down the following inequality
\begin{equation}\label{bound0}
\gamma m\hbar\omega^2\leq \frac{4\omega}{30}
\end{equation}
where we have used the final transition rate expression as it shall provide us with the tightest bound on the $\gamma$ parameter. The most sensitive frequencies of resonant bar detectors are of the order of $1$ kHz. Events like collapsing and bouncing cores of supernovas can produce gravitational waves in the vicinity of $1-3$ kHz. Hence, for resonance to occur the frequency of the harmonic oscillator must lie in the range of $1-3$ kHz. On the other hand, the mass of such a resonant bar detector is of the order of one ton ($1.1\times 10^3$ kg). Instead of using $\omega\sim 1$ kHz, we use the maximum value of $\omega$ needed to detect gravitational wave with maximum frequency $3$ kHz. Hence, $\omega=3$ kHz. Now it is important to note that the `L' shaped interferometer detectors are more efficient than the general resonant bar detectors presently in existence. In order to increase the sensitivity of the detectors, one needs to use more sophisticated technology which will provide us with a higher gain in the change in the total length while being at the resonance condition. Such a technology may be useful in the future when bar detectors can be made much longer in length than the ones currently present. For the lower sensitivity of the bar detectors, the bounds on the GUP parameters become indeed weaker but if a much more sensitive detector is made it will be possible to predict tighter bounds on the same. Currently, few attempts are being made to increase the sensitivity of the bar detectors to $10^{-21}$ by cooling them down to milli-Kelvin temperature and using the full capability of the bar detectors \cite{Pizzella}. Resonant bar detectors with spherical geometry are also being considered for substantially increasing the sensitivity of the bar detectors \cite{CostaAguiar}.

\noindent Now, $\alpha=\frac{\alpha_0}{m_p c}$ and $\beta=\frac{\beta_0}{m_p^2c^2}$ with $m_p\sim2.176\times10^{-8}$ kg denoting the Planck mass.  We can now express $\gamma$ in terms of a dimensionless parameter $\gamma_0$ as
\begin{equation}\label{gamma}
\begin{split}
\gamma&=\left(\frac{\alpha^2}{2}+\beta\right)=\frac{1}{m_p^2c^2}\left(\frac{\alpha_0^2}{2}+\beta_0\right)\\
&=\frac{\gamma_0}{m_p^2c^2}
\end{split}
\end{equation}
where $\gamma_0=\frac{\alpha_0^2}{2}+\beta_0$. In terms of $\gamma_0$, eq.(\ref{bound0}) can be recast in the following form
\begin{equation}\label{bound1}
\begin{split}
\gamma_0&\leq \frac{2  m_p^2 c^2}{15\hbar m\omega}\\
\implies \gamma_0&\leq 1.6\times 10^{28}~.
\end{split} 
\end{equation}
We now need to obtain bounds on the dimensionless parameters $\alpha_0$ and $\beta_0$ individually. In order to do so, we need to use the inequality obtained earlier in \cite{sbprd} while using the path integral formulation of a point particle moving in an arbitrary potential. The inequality obtained in \cite{sbprd} reads
\begin{equation}\label{GUPrel}
\beta>4\alpha^2
\end{equation}
which in terms of the dimensionless parameters read $\beta_0>4\alpha_0^2$. using this relation back in eq.(\ref{bound1}) we obtain the bounds on the dimensionless GUP parameters as $\alpha_0\leq 10^{14}$ and $\beta_0\leq 10^{28}$. The bound on the $\beta_0$ parameter is at par with the bound obtained in \cite{sbf} but tighter than the bound obtained in a graviton-detector interaction case \cite{OTMgraviton} in a quadratic GUP framework. We shall now make comparisons among the bounds obtained in our paper with some of the bounds currently in existence among other literature. For example, in \cite{BigBangNucleosynthesis} using observational bounds from the Big Bang nucleosynthesis and primordial abundances of some of the light elements, bound on the quadratic GUP parameter was obtained. In \cite{BigBangNucleosynthesis}, the most stringent bound that was obtained for $\beta_0$ was $\beta_0\leq 10^{81}$ which is much weaker than the bound obtained for our case ($\beta_0$ was $\beta_0\leq 10^{28}$). In \cite{LorentzViolationGUP}, a procedure was proposed to connect the $\beta_0$ parameter to coefficients of the gravity sector in the standard model extension where violation of Lorentz invariance is encoded into specific parameters. The bound obtained in this process was $\beta_0\leq 10^{51}$ which is again way weaker than the bound obtained in this analysis. In \cite{BaryonAsymmetryGUP}, the Baryon asymmetry problem was explained using the GUP framework which gave the values of the two GUP parameters to be $\alpha_0\approx 10^4$ and $\beta_0\approx -10^{8}$, which are way tighter than the bounds obtained in our current analysis.

\noindent Another insight into the entire problem comes from the viewpoint that the position and momentum observables that we work with correspond to the center of mass (COM) motion of the resonant bar detector.
If the GUP modified phase space variables of the COM of the bar detector are $\hat{Q}_{\mathcal{C}}$ and $\hat{P}_\mathcal{C}$,  then the modified commutation relation in a one-dimensional model has some differences from that of the commutation relation given in eq.(\ref{gupu}). This has been explicitly analyzed in the literature \cite{Amelino} and later a detailed discussion regarding the relativistic and field theoretical generalization has been done in \cite{Bosso3}. In \cite{Bosso3}, a critical discussion of the ongoing GUP framework has been done and its application to composite systems has been thoroughly investigated.

\noindent In \cite{Amelino}, for a macroscopic body with $N$ constituent particles, the commutator between $\hat{Q}_\mathcal{C}$ and $\hat{P}_\mathcal{C}$ is given by (in one dimension)
\begin{equation}\label{5D.1}
[\hat{Q}_\mathcal{C},\hat{P}_\mathcal{C}]=i\hbar\left(1-\frac{\lambda'}{N}\hat{P}_\mathcal{C}+\frac{\lambda^2}{N^2}\hat{P}_\mathcal{C}^2\right).
\end{equation}
Comparing the parameter appearing in the above equation with our results, we observe that the quadratic and linear GUP parameters, $\alpha$ and $\beta$ scale as
\begin{equation}\label{5D.2}
\alpha\sim \frac{\alpha_0}{Nm_pc}, ~\beta\sim\frac{\beta_0}{N^2m_p^2c^2}~. 
\end{equation}
As a result $\gamma$ shall also scale as $\gamma\sim\frac{\gamma_0}{N^2m_p^2c^2}$. Now if we consider one mole of constituents per gram of the bar detector, then one gram of the bar detector will consist of Avogadro number of particles ($6.023\times 10^{23}$). Hence for a $10^3$ kg bar detector, there are approximately $10^{29}$ number of particles. As a result the bound on  $\gamma_0$ becomes $\gamma_0\leq \frac{2m_p^2c^2}{15N^2\hbar m\omega}\sim 10^{86}$. Hence, we obtain $\beta_0\leq 10^{86}$ and $\alpha_0\leq 10^{43}$. These bounds obtained using COM observables are not very stringent. 
What is important to note for a gravitational wave fluctuation, the center of mass is not displaced rather only the bar is stretched and squeezed and hence, the bounds obtained using the COM degrees of freedom are redundant. We have included this analysis only for the sake of completeness of our current calculation. What we have looked at in the discussion earlier can be thought of as variables not corresponding to the COM variables and which is applicable in our current analysis. For the response of the detector to the gravitational wave, one needs to look at the collective microscopic vibration of individual perturbations and hence the bounds obtained are strictly $\alpha_0\leq 10^{14}$ and $\beta_0\leq 10^{28}$.

\section{Conclusion}
\noindent In this paper, we explore the quantum responses of the resonant bar detectors to the gravitational waves in the presence of quantum gravity corrections in the Heisenberg uncertainty principle. Here we have derived the modified quantum states of a bar detector and the transitions between them induced by the incoming gravitational wave in the extended generalized uncertainty principle framework, namely the LQGUP framework. This relation contains both the corrections linear and quadratic terms in momentum. Firstly, the existence of the linear and quadratic generalized uncertainty principle makes a shift in the non-degenerate states of the one-dimensional harmonic oscillator. The energy eigenvalues also get modified by the LQGUP parameters $\alpha $ and $\beta$. Then we derive the observable effects in the transition rates from the ground state to the excited states to trace out any possible relic effects of the generalized uncertainty principle in the gravitational wave data from the resonant bar detectors. From the exact forms of the transition rates, we have made the following observations. 
 
\noindent In the presence of the linear and quadratic generalized uncertainty principle, there are four possible transitions from the modified ground state $|0^\beta \rangle$ to the higher excited states up to $|4^\beta \rangle$. Eq.(\ref{10}) together with eq.(\ref{conts}) shows the transition between $|0^\beta \rangle\rightarrow |1^\beta\rangle$ and $|0^\beta \rangle\rightarrow |3^\beta\rangle$ which occur only due to the linear GUP parameter $\alpha$. Therefore, these two transitions ensure the contribution of the linear order correction in momentum in the modified uncertainty principle which is absent for the ordinary HUP and quadratic GUP structures. Again the transitions $|0^\beta \rangle\rightarrow |2^\beta\rangle$ and $|0^\beta \rangle\rightarrow |4^\beta\rangle$ bear the signature for both $\alpha$ and $\beta$. Note that in the limit $\alpha, \beta \rightarrow 0$, there is only one transition due to $T_{02}= \frac{1}{2\sqrt{2}}$. Thus resonant detectors of the gravitational wave are not only a reliable candidate to probe such GUP effects but also able to predict the mathematical structure of such Planckian modifications. Now we have the resonant frequencies $\Omega=\omega(1+3\gamma m \omega \hbar)$, $\Omega=\omega(2+9\gamma m \omega \hbar)$, $\Omega=\omega(3+18\gamma m \omega \hbar)$  and $\Omega=\omega(4+30\gamma m \omega \hbar)$, where $\gamma= \left(\frac{\alpha^2}{2}+\beta\right)$. Hence the frequencies at which the resonant detector responds to the incoming gravitational waves get modified by the generalized uncertainty principle parameters $\alpha$ and $\beta$. We hope that such effects in the resonant frequencies will be detectable in the near future if the generalized uncertainty principle exists. This observation is quite similar to that of the noncommutative structure of space \cite{sg4}, \cite{sg5}.   The transition rates $P_{0^\beta \rightarrow 1^\beta}$ and $P_{0^\beta \rightarrow 3^\beta}$ contain the quadratic term in the dimensionless parameter $\alpha$ which is of the linear order of $\beta$. Again both the linear and quadratic terms in $\beta$ appear in the transition rates. The linear dependence in $\beta$ is easier to detect.   Our analysis shows that both linear and circularly polarized gravitational waves are good candidates to probe the presence of the generalized uncertainty principle in the resonant detectors. This observation is valid for both the periodic and aperiodic signals as well. Finally, we obtain bounds on the dimensionless GUP parameters $\alpha_0$ and $\beta_0$. We observe that the bound on the $\beta_0$ parameter agrees with the result obtained earlier in the case of a similar setup with only quadratic GUP framework \cite{sbf}. The observations made in this paper reveal that resonant detectors may allow in the near future to detect the existence of an underlying generalized uncertainty principle framework. Moreover, these gravitational wave data in the resonant detectors can also be helpful in constructing the mathematical form of the proposed quantum gravity effect in the generalized uncertainty principle.

\end{document}